# Policy Network Approach to Coordinated Disaster Response

Kwang Deok Kim and Liaquat Hossain

*Abstract—* **In this paper, we explore the formation of network relationships among disaster relief agencies during the process of responding to an unexpected event. The relationship is investigated through variables derived from the policy network theory, and four cases from three developed countries such as (i) Hurricane Katrina in the US; (ii) Typhoon Maemi in South Korea; (iii) Kobe; and, (iv) Tohoku Earthquake in Japan that failed to cope with extreme events forms the basis for case study presented here. We argue that structural characteristics of multi-jurisdictional coordination may facilitate or impede in responding to a complex nature of today's disaster. We further highlight the promise of policy network approach in facilitating the development of multi-jurisdictional coordination process which may provide new avenue to improve the communication and coordination of hierarchical command control driven organizations with the local community. Our proposed novel approach in investigating the usefulness of network approach through media content analysis for emergency may provide opportunity as a countermeasure to a traditional hierarchical coordination, which may give further insights in establishing a more effective network for emergency.**

## I. INTRODUCTION

Research in disaster management puts primary effort on the government organizations due to the availability of extensive resources, physical forcing, and the legal authority, which may contend with disasters. In disaster response, however, the government sector alone cannot respond effectively and timely during emergency. The changes in social structure, the role change of government organizations, and diverse aspects of a disaster cause the limitations. As a result, the private sector's participation, such as NGOs and private organizations as well as community based volunteer organizations, has been seen increasingly important in disaster response management. Furthermore, the transition to the network approach is often raised to be needed due to the nature of a disaster management system in the traditional bureaucratic approach. "A long tradition of bureaucratic approach is slowly surrendering" as Taylor [1] addresses, and a network approach [2, 3] is suggested as a possible breakthrough as a means to tackle problems in disaster management [4].

In particular, we utilize a policy network approach to examine the dynamic features of responding participants and policy making inclusive of policy implementation during a disaster response phase. For the initial exploration, network variables are drawn from the policy network theory to understand the features of *participants* with a focus on *relationships* between them. These variables are examined with media content analysis.

This paper begins with a research question, "*How does a policy network perform in the process of a disaster response?*" We look for the difference from network components (network variables in details) in policy network theory and network types following disaster response management systems to tackle the question. The following questions are derived from the research question:

1. Who are the participating actors for a disaster management system during a disaster response?
2. Who does play a central role to coordinate activities among participating actors?
3. What are the relationships in the network among participating actors?
4. Is the network opened or closed? and, What is or are the inherent benefits of open, close or a hybrid coordinating structure in providing effective and timely response?

The structure of the paper is as follows: Section 2 describes a policy network approach with types and components, and presents a conceptual framework; Section 3 observes a research approach in disaster management. A framework of case study and data collection are examined; Section 4 investigates network variables for each case, i.e., Katrina, Maemi, Kobe, and Tohoku; Section 5 presents implications of the analysis results and recommends a desirable network model for emergency.

## II. POLICY NETWORK THEORY

The term policy network is used to examine and analyze the behaviours and interactions among actors (i.e., participants) [5]. Policy network has won recognition of analyzing structural relations among participants in policy process [6]. Under rigid hierarchy, it is difficult to take an appropriate response to a fast-paced environment, and researchers who experienced the limitations from hierarchical bureaucratic government pay attention to policy network as a new approach. Yet, empirical studies on policy network that examine the behaviours and interactions of participants in disaster management are rarely found [5, 7, 8].

Rhodes [9] suggests that the policy network as a comprehensive concept, which is arranged on the various types of network and classifies the policy networks into five kinds: (i) *policy/territorial community* with the closed form of the type of policy networks; (ii) *professional network* with professional groups exercising a dominant influence; (iii) *inter-governmental network* with the central government and

Kwang Deok Kim is with the Faculty of Engineering and Information Technologies, the University of Sydney, NSW 2006, Australia (phone: +61 2 9351 2247; e-mail: kwang.kim@sydney.edu.au).

Liaquat Hossain is with Information Management Division of Information and Technology Studies, the University of Hong Kong and also Honorary Professor, Complex Systems School of Civil Engineering Faculty of Engineering and IT, the University of Sydney, Australia (e-mail: lhossain@hku.hk, liaquat.hossain@sydney.edu.au).

local governments regarding the relations; (iv) *producer network* with the economic group having a dominant influence; and, (v) *issue network* with more openness. These are classified "ranging along a continuum from highly integrated policy communities to loosely integrated issue networks" and "according to their membership composition, the extent of interdependence between their members, and the distribution of resources between members." [10]

TABLE I. COMPONENTS AND VARIABLES OF POLICY NETWORK FOR THE CURRENT STUDY

| Components | | Variables | Description |
|---|---|---|---|
| Participants | | Actors | The government sector, the private sector |
| | | Attitude | Is the behaviour active or passive? |
| | | Motivation | Existence of incentive |
| Relationships | Interaction | Frequency | Whether relationship is formed and whether participants have a contact, such as a meeting |
| | | Direction | One-sided or mutual |
| | Network Structure | Interdependence | Cooperative or conflictive |
| | | Exclusion | Is the network open or closed to participants? |

Recent studies tend to analyze policy network with adding or subtracting of the components developed from the Rhodes' model and reported in Marsh and Rhodes [11] [5, 10]. Furthermore, to understand the dynamics of policy network, it is helpful to explore definition of the term 'policy network' is. Montpetit [6] states and Lee and Park [5] rephrase the term, the policy network refers to the *relationships* among *participants (or actors)* through the policy making process. Thus, in this study, these two concepts are centered and considered for the selection of components of policy network with Rhodes' model [11] as a basic foundation.

As for *participants*, Lee and Park [5] states that participants are groups or individuals who take part in the policy making process to accomplish the common goal. This component, participants, has three subcomponents which can be measured by the number of participants and the nature of the participants, such as attitudes and motivation for participation [5, 12]. Israel and Rounds [13] highlights the size of number of participants who involved in the network to measure as a component for analysis. The component is examined from the perspective of the group unit in this study.

As for *relationships*, interaction and a type of relationships are needed to be considered to understand the notion of the policy network [5]. Firstly, interaction means a substantial process that resources are exchanged and mobilized to make the common goal real [14]. To measure this component interaction, the frequency of interactions and the direction of relationships are used and examined as subcomponents in the policy network [5]. Mitchell [15] and Israel and Rounds [13] describe the interaction as an analysis component to examine the frequency and the reciprocity among participants. Secondly, a type of relationships refers to the network structure between participants which has an influence on outcomes [16, 17]. Yishai [18] addresses the five types of policy network are distinguished by two variables [1] ; interdependence relating to relationships between participants, and exclusion pertaining to the degree of openness to new participants. To measure this component network structure, two variables are suggested in Lee and Park [5]: (i) whether relationships between participants are cooperative or conflictive; and, (ii) whether it is easy for new participants to take part in and withdraw from a network structure. Kim argues [19] that the basic element of the network is the relationship among social components because the network changes by the relationships that actors form. In addition, Mitchell [15] considers the network structure as a component for analysis. TABLE I shows the components and variables of policy network considered in previous literature and used for analysis in this study.

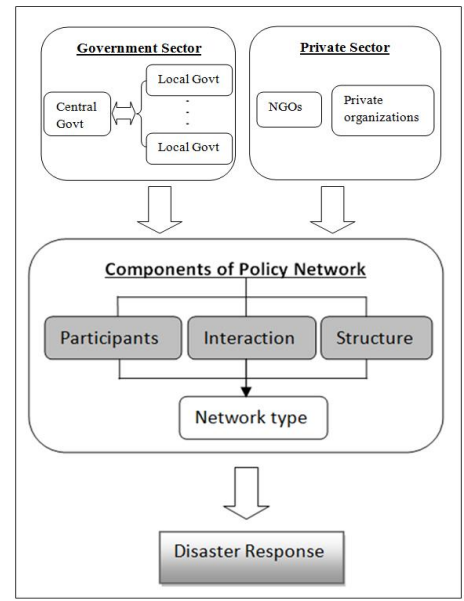

Figure 1. A conceptual framework

In order to explore the dynamics of policy network, it is important to investigate the components constituting the policy network. As for the selection of the components, this study follows the same route as the previous studies. As observed from above, there are various components used in previous literature to conduct an analysis and it is found that they tend to form the components from the basis of Rhode's model [3] as Lee and Park [5] addresses in their recent work on policy networks in disaster management. The commonly used components of policy network proposed from the recent

---

[1] In this paper, the term "subcomponent" is replaced with the term "variable" as used in [18] to measure the components of the policy network for analysis.

studies are chosen for this study. Fig. 1 shows the conceptual framework for this study.

Disaster management in each country has been formed differently in terms of an institution. Due to these different institutional backgrounds, we argue that components of policy network may work differently, and that a different type of policy network may be formed under the influence of them. The type of policy network may have an effect on a difference appearing in responding to a disaster. A desirable network in disaster management responding effectively to disasters will be discussed. By comparing the four cases in three countries, the difference of network components and network types will be driven. A research approach in disaster management is observed in the following section.

### III. RESEARCH APPROACH

It is important to understand why the networks differ in various contexts during disaster response. In order to examine this process of network formation among participants, three countries among well-developed countries in terms of disaster management that failed to cope with the latest extreme emergencies are selected for this paper [20, 21]. In order to compare disasters in the U.S., Korea, and Japan, the following cases are chosen for the study: Hurricane Katrina in the U.S. in 2005, Typhoon Maemi in Korea in 2003, the Kobe earthquake in the southern part of Hyogo Prefecture, Japan 1995, and the Tohoku earthquake in Japan in 2011.

Content analysis has been used in various fields, such as sociology, communication, and journalism for more than five decades [22]. In particular, researchers in sociology, such as Max Weber, have had interests in media content since the early 20th century to measure the "cultural temperature of society" [23]. Media content analysis has become gradually prevalent as a research method during the 1920s and the 1930s [24]. Afterwards, it has been primarily used as a quantitative research method which coded text of content data into clear categories, and then described by using statistics [25]. In order to code a text or set of texts during content analysis, there are a number of choices (or eight category coding steps) needed to be made, such as level of analysis and generalization. These choices are important not only these are being made due to "potential methodological bias but also because such choices require semantic interpretations of the data" [26]. These steps affect results received and the interpretation of results adapted from Carley [27]. TABLE II shows codes for content analysis. There are seven themes for three categories.

In the category of participants: first, the theme *actors* means that how many organizations take part and which organizations play a principal role in a response phase; second, *attitude* implies whether the behaviour of responding organizations is active or passive; third, *motivation* refers to whether there is any incentive to encourage organizations to coordinate and participate.

In the category of interaction: first, *frequency* is related to how participating organizations communicate or have a contact to exchange information vital for coordinating response actions; second, *direction* means that whether the exchange of resource, such as personnel and equipment among participants is one-sided or dyadic.

In the category of network structure: first, *interdependence* refers to whether the relationships between participants is cooperative or conflictive; second, *exclusion* means whether the access of new participants to resources for disaster relief actions is easy or whether the access of news participants to the network structure is easy to participate in or withdraw from.

TABLE II. CODES DEVELOPED FOR CONTENT ANALYSIS

| Category | Theme | Code | Sub Code |
|---|---|---|---|
| **Participants** | Actors | ACTS | ACTS_CTGOV<br>ACTS_LCGOV<br>ACTS_PRV |
| | Attitude | ATTD | ATTD_ACT<br>ATTD_PAS |
| | Motivation | MTVN | MTVN_POS<br>MTVN_NEG |
| **Interaction** | Frequency | FREQ | N/A |
| | Direction | DRCN | DRCN_ONE<br>DRCN_DYD |
| **Network Structure** | Interdependence | INTD | INTD_COOP<br>INTD_CONF |
| | Exclusion | EXCL | EXCL_OPN<br>EXCL_CLS |

Notes: ACTS_CTGOV(Actors_CentralGovt), _LCGOV(LocalGovt), _PRV(PrivateSector), ATTD_ACT(Attitude_Active), _PAS(Passive), MTVN_POS(Motivation_Positive), _NEG(Negative), DRCN_ONE(Direction_Onesided), _DYD(Dyadic), INTD_COOP(Interdependence_Cooperative), _CONF(Conflictive), EXCL_OPEN(Exclusion_Open), _CLS(Closed)

Coding category in TABLE II reflects interpretations of disaster management in responding to an extreme event. An initial pilot run for a coder training and a test of reliability was conducted. The coding instruments and procedures were sophisticated as well through the initial run. In particular, to establish reliability of coding, a coder was asked to conduct coding for sample contexts excerpted randomly. Afterwards, the both results were examined for the portions of the coding and agreed with little ambiguity. In order to reduce bias in the study and to keep it objective, a senior researcher who was not familiar with the four cases was involved in early analysis.

This paper uses various media data from The New York Times, Associated Press, Cable News Network (CNN), Munwha Broadcasting Cooperation (MBC), Korean Broadcasting System (KBS), Seoul Broadcasting System (SBS), Chosun Ilbo, Yonhap Television News (YTN), Donga Ilbo, Nihon Hoso Kyokai (NHK), Fuji TV, TV Asahi, Washington Post, Japanese Weekly Magazine, The Sankei Shimbun. Number of articles identified for the analysis in each case: 18 for Katrina; 16 for Maemi; 17 for Kobe; 20 for Tohoku.

## IV. MEDIA CONTENT ANALYSIS

For each case, this study examines network components derived from the reviews of literature, which enables to understand [1] who are the participating actors for a disaster management systems during a disaster response? and, who plays a key role to coordinate activities among actors?, [2] what are the relationships in the network among actors? [3] is the network opened or closed? and, *What is or are the inherent benefits of open, close or a hybrid coordinating structure in providing effective and timely response?*

### A. Hurricane Katrina

Hurricane Katrina was considered just as one of the tropical storms at the beginning on August 23, 2005 [28]. This tropical storm, however, grew into a disastrous hurricane over the next seven days. All the citizens in the US did not think the tropical storm seriously and watched with a curiosity, but soon it turned into deep concern. American citizens held the awes for the unbelievable ferocity of the catastrophic event and became disappointed and frustrated at the incompetency of the government sector to respond effectively [29]. Even though Hurricane Katrina was one of the most predictable natural disaster in US history, the authorities failed to manage to respond effectively to the crisis [30, 31]. In due consideration of hurricane occurrences and accuracy in predictions, this was not common in the US [32]. Katrina and the following flooding disclosed momentous defects in the government sector – Federal, State, and local – for responding to catastrophic events [29]. Hurricane Katrina showed the world how susceptible the US can be to natural disasters and it was recorded as one of the most expensive natural disasters in US history [33].

### B. Typhoon Maemi

Super Typhoon Maemi, the strongest one ever since records began nearly 100 years in Korea, hit the southern coastal region with a category 4 on the 12th of September in 2003. The typhoon ravaged the south part of Korea with the atrocious wind reaching up to 216 kph and 450 mm of precipitation, which consequently caused the tragic loss of more than 117 lives and 25,000 people homeless [34]. Since the track of Typhoon Maemi was closer to industrial areas, it brought huge damage in industrial infrastructure as a result. Even though there was harsh criticism of the slow disaster response from the Korean government during and after Rusa in 2002, the authorities did not seem to learn any lessons from their past mistakes. Furthermore, they did not carry out effective disaster relief measures, such as evacuation operations, when Maemi made landfall one year later.

### C. Kobe Earthquake

The Great Hanshin Awaji earthquake commonly referred to as the Kobe earthquake because it struck the Kobe region (henceforth called the Kobe earthquake), shook Japan at 5:46am on 17 January 1995 with a magnitude 7.2 on the Richter scale and a depth of 13.2km. The strong quake ravaged densely populated regions, leaving many casualties and immense property damage [35]. Even though Japan often experienced many earthquakes with a various size of a magnitude and was well prepared, it was a great shock even to the well experienced country with this size of a magnitude in 1995 [36]. There was no such a destructive quake which shook Japan in the 20th century except for the great Kanto earthquake that devastated wide regions of Tokyo and Yokohama, leaving 143,000 casualties in 1923 [37].

### D. Tohoku Earthquake

The East Japan Great Earthquake or The 2011 off the Pacific coast of Tohoku Earthquake (for simplicity, henceforth called the Tohoku earthquake) named by the Japan Meteorological Agency occurred off the Sanriku coast of the Tohoku region at 2:46 pm. (Japan Standard Time) on 11 March 2011 with a magnitude 9.0 on the Richter scale [38]. A massive tsunami over 10m high induced by the quake struck the north-eastern part of Japan. In addition to the initial colossal earthquake and the subsequent tsunami, the nuclear power stations in Fukushima Prefecture which were the major source of electrical energy for the Kanto region were seriously damaged in consequence of the massive hit from the tsunami [39-41]. In particular, the threat coming from the consequences of the meltdown is still real and ongoing more than two years after the horrifying incident as of May 2013 [42]. The triple disaster left 15,797 people died with 3,054 missing as of April 2012 and caused the estimated damage amounted to the tens of billions US dollars [40, 43].

### E. Frequency of Codes

TABLE III. FREQUENCY OF CODES IN MEDIA SOURCES

| (sub) Codes | Katrina | | Maemi | | Kobe | | Tohoku | |
|---|---|---|---|---|---|---|---|---|
| | Freq | % | Freq | % | Freq | % | Freq | % |
| ACT_CTGOV | 68 | 61 | 44 | 52 | 72 | 41 | 92 | 58 |
| ACT_LCGOV | 34 | 30 | 18 | 21 | 64 | 36 | 34 | 21 |
| ACTS_PRV | 10 | 9 | 22 | 26 | 40 | 23 | 34 | 21 |
| ATTD_ACT | 18 | 39 | 12 | 31 | 4 | 20 | 6 | 20 |
| ATTD_PAS | 28 | 61 | 26 | 68 | 16 | 80 | 24 | 80 |
| MTVN_POS | 8 | 10 | 2 | 12 | 8 | 10 | 20 | 17 |
| MTVN_NEG | 76 | 90 | 14 | 88 | 72 | 90 | 96 | 83 |
| FREQ | 24 | n/a | 2 | n/a | 16 | n/a | 6 | n/a |
| DRCN_ONE | 18 | 90 | 6 | 75 | 20 | 71 | 30 | 88 |
| DRCN_DYD | 2 | 10 | 2 | 25 | 8 | 29 | 4 | 12 |
| INTD_COOP | 4 | 15 | 2 | 50 | 16 | 67 | 6 | 33 |
| INTD_CONF | 22 | 85 | 2 | 50 | 8 | 33 | 12 | 67 |
| EXCL_OPN | 6 | 23 | 2 | 25 | 0 | 0 | 14 | 100 |
| EXCL_CLS | 20 | 77 | 6 | 75 | 12 | 100 | 0 | 0 |

Content analysis is done manually and every data source is examined line by line. Rather than just presenting the frequency of words within a text, this study focuses on each concept for content analysis as a more sophisticated analysis as Pool [44] underlines in the book titled, 'Trends in content analysis'. The frequency of a certain concept in a text is counted and calculated on the basis of its code in the end.

Concepts and codes developed in the previous section are used for content analysis.

TABLE III presents the frequency and percentage of the codes (sub codes) that are counted and accumulated one by one through examining the media sources. Each percentage is only applied among codes of the same theme. As for percentage of *Actors*, for example, 61% for ACTS_CTGOV of Katrina in the following table means that it takes up 61% out of the entire sum for the theme *Actors*.

*F. Content Anaysis*

In this paper, we utilized the policy network to examine the dynamic features of responding participants and policy making inclusive of policy implementation. Network variables were drawn from the policy network theory to understand the features of *participants* with a focus on *relationships* between them. These variables were examined with media content analysis. In this section, we compare the four disaster cases and examine the findings discovered in the previous section 4 by looking at data in divergent ways.

In particular, two approaches are utilized to go beyond the initial impressions and derive a new concept by looking for each case's similarities with intergroup differences. A first approach suggested by Eisenhardt [45] is to search for patterns. This approach selects categories first, and then looks for within-group similarities with intergroup differences afterwards. A second approach is for selecting pairs of cases, and then for listing the similarities and differences between each pair. This approach helps investigators look for the delicate similarities and differences between cases. That is to say, a search for similarities from a seemingly different pair can provide researchers with more matured understanding and in the same manner, a search for differences from seemingly similar cases by researchers can give an opportunity to break simplistic frames for sophisticated understanding. The result gained after these comparisons can be used as new categories and concepts that researchers never expected beforehand. Grouping cases into threes or fours for comparison is an extension of this approach [45].

*A. Approach One*

With the first approach, we particularly examine the relationships among joining participants with the central government centered on as shown in Fig. 2.

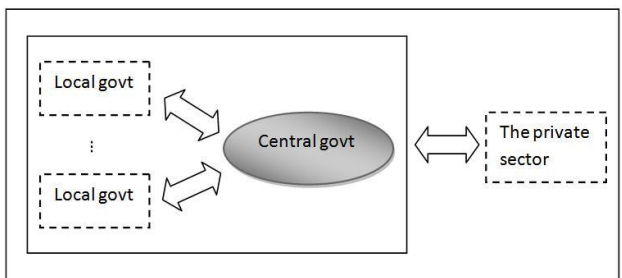

Figure 2. Approach One:
Relationships with the central government being centered

For this approach, four themes in each case are selected; actors, attitude, direction, and exclusion. As for the code *actors*, the number of participants interacting in response operations and which organizations respond first are observed. For *attitude*, whether central governments (e.g., the federal government in the US) support the first responders active to respond effectively to the crisis is examined. As scholars [46, 47] argue that all emergency management systems today are modelled on a bureaucratic approach with each case may have one-sided relationships with outer actors (or new actors joining in) in common in terms of sharing information and resources. Thus, *direction* and *exclusion* from two categories – interaction and network structure - are considered for investigating whether the direction of resource is one-sided or dyadic, and whether network structure is opened or closed.

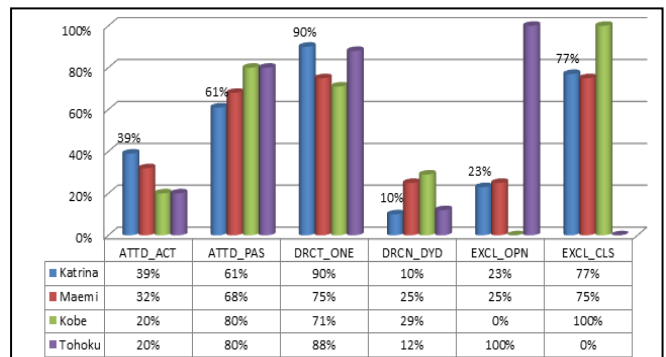

Figure 3. Vertical cylinder graphs presenting the proportions of the selected categories for the first approach of cross-case analysis

As depicted in Fig. 3, all four cases have similarity and difference each other; (1) it is clear that there is a similar pattern found in two concepts, i.e. *Attitude* and *Direction*, (2) and from the last concept, i.e. *Exclusion*, there is a different pattern shown, in particular, from the two cases, Kobe and Tohoku.

As for a similar pattern, two concepts appeared in having a similarity for four cases. First, while the proportion of an active attitude among responding participants for each case is accounting for a relatively small ratio, between 20% and 39%, the passive attitude of participants in an unpredicted event has a much higher proportion, from 61% to 80%. Most of passive attitude was from the government sector and attitude of the private sector was active and swift in action in most cases. In particular, it is interesting that the results of the two Japanese cases for the active attitude and the passive attitude are exactly the same, 20% for active and 80% for passive. It is obvious that there is no change at least in the attitude among participants between the 1995 Kobe earthquake and the 2011 Tohoku earthquake. Second, the proportion of the one-sided direction in exchanging information inclusive of resource is accounting for a great amount, from 71% to 90%, in marked contrast to the proportion, between 10% and 29%, of the dyadic direction in information exchange. This result presents that each disaster response system has a strong tendency in conveying information and resource one sided rather than dyadic between the responding participants, which proves that all disaster response systems today are modelled on a bureaucratic approach with one-way relationships in terms of sharing information and resource, as scholars argue [46, 47]. As for the proportion of direction in Japanese cases, it is interesting that the one-sided direction ratio increased by 17% from 71% for the Kobe case to 88% for the Tohoku case, after considering all of those reports praising the lessons the Japanese government learned since the Kobe temblor. This may be explained by the authorities trying to conceal or downplay the facts of the severity of the nuclear power plants in. Rather than telling the truth and sharing information, the government was busy hiding the critical information, such as meltdown in the plants and the level of evacuation.

The first two cases from concept *exclusion*, account for the almost same proportion, 23% and 25% for a closed network, and 77% and 75% for an opened network. However, for two Japanese cases, there is an ultimate difference in each concept, between 0% and 100%. This is the most notable difference among other concepts and proposes a big change in a network structure between the government sector and the private sector showing Japanese authorities have the possibility of learning in a disaster response system which can respond in an effective manner.

### B. Approach Two

As for the second approach, four cases are separated into two groups and this classification is made by whether disaster management systems considering local governments as first responders with help from the private sector are established. Approach two analysis focuses more on two paired cases with similarities and differences.

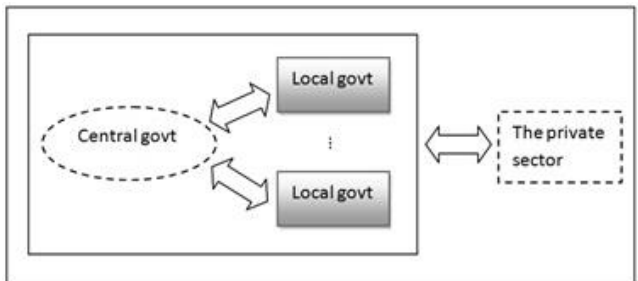

Figure 4. Approach Two:
Relationships with the local governments being centered

As for grouping, Japanese cases are divided into two different groups. The U.S. and Japan are evaluated as a country where local governments take the first responsibility to respond and cope with central government and the private sector [29, 48]. In case of Kobe, however, rather negative evaluations of government performance have centered around delays in the mobilization of critical resources and the initiation of key response tasks [49]. In case of Maemi, Korea did not have disaster management systems with the adjusted roles between central government and local governments to cope with the disaster effectively [50]. Thus, the Kobe case is coupled with the case of Maemi and the other two cases – Katrina and Tohoku - form a group.

While the group 1 consists of Katrina and Tohoku, group 2 is composed of Maemi and Kobe. The classification is based on whether a disaster management system urging the importance of the local government's role has been set up prior to a certain disaster or not. Firstly, as for the Katrina case, the U.S. had an integrated management system based on FEMA and enacted laws for local governments to respond first in case of emergency and other government agencies were supposed to support them if a scale of a disaster is large. Secondly, for the case of Maemi, it was arranged for local governments to respond first, but it was not long after the local self-government system was put into practice and the system was not settled, which caused an ambiguous attitude in taking clear responsibility among the central and local governments. Thirdly, for the Kobe case, while there was an earthquake resistance design code introduced in 1981, Japan had no true comprehensive disaster management plan in coordination with local governments as practical first responder and voluntary organizations. Lastly, on the basis of the lessons learned from the Kobe case, the Japanese government focused on a unified disaster management with local governments as first responder and utilizing voluntary efforts in a better coordinated way. With the second approach of the cross case analysis, this study examines the relationship among participants with the local government centered around. Furthermore, the study also investigates if there is a relation between the type of disaster management system and the effectiveness of the system during a disaster response phase. The following Fig. 5 presents the average proportions of each code based on groups. For instance, the proportion 60% of ACTS_CTGOV in the group 1 came from the average of 61% in the same code of Katrina and 58% of Tohoku.

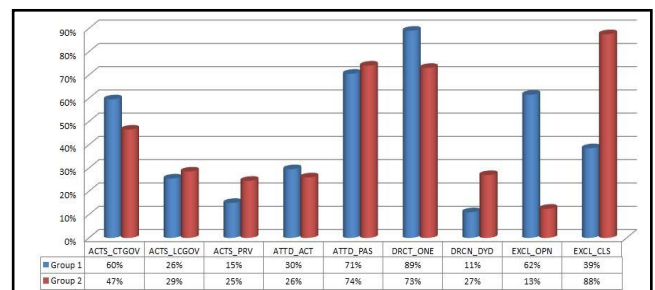

Figure 5. Vertical cylinder graphs showing the average of proportions for each concept by group

Firstly, for the concept *actors*, it is obvious that the management systems established previously have no effect on the disaster response with the local government centered around coordination with the private sector. If the systems of group 1 were put into practice efficiently as planned, the proportion of the local government and private sector would have been taken up more than the proportion of group 2, which suggest lack of effectiveness of the management systems by group 1. Secondly, as for concepts *attitude* and *direction*, it is hard to find any efficiency of utilizing the management systems by group 1. Even though there is a slight improvement in the active attitude by 4%, it is hard to say that there is a significant difference after setting up the management systems. Furthermore, in the direction concept, the ratio of the dyadic direction in the group 2 is larger than the one of group 1 by 16%. Lastly, for the concept *exclusion*, there is a significant difference between group 1 and group 2. While the proportion of the opened network of group 1 is 62%, only 13% is made up of group 2. It is clear that the disaster management systems of group 1 have an influence on the network structure and helped the government sector flexible to accept relief efforts from outside, such as NGOs.

## V. CONCLUSION

### A. Implications of Findings

In this study we seek to answer question *how does a policy network perform in the process of a disaster response?*, which has been carried out looking for the difference from network components (network variables in details) in policy network theory and network types following disaster response management systems to tackle the question. The following TABLE IV presents the results based on the previous analyses for the questions raised from the previous section 1.

TABLE IV. RESULTS FOR THE RESEARCH QUESTIONS

| Case | Q1 | Q2 | Q3 | Q4 |
|---|---|---|---|---|
| **Katrina** | The government sector & the private sector | Local govt. was supposed to but overwhelmed | One-sided and lacked of communication | Closed for helps from outside, but opened slightly through NGOs |
| **Maemi** | The government sector & the private sector | Varied according to local govt. | One-side, but dyadic for some local govt. | Closed, but opened slightly through NGOs |
| **Kobe** | The government sector & the private sector | Local govt. was expected to but overwhelmed | One-sided and swamped with the sheer number of volunteers | Closed tight |
| **Tohoku** | The government sector & the private sector & international relief organizations | Local govt. and emergency headquarter, but lacked of coordinated disaster management framework | One-sided and did not have a clear command line due to a lack of leadership at the political level | Opened widely for the outside help |

The policy network of disaster response management for the three cases, i.e. Katrina, Maemi, and Kobe, were formed of the most closed network, a *policy community* type out of five types of policy networks classified by Rhodes and Marsh [10]. On the other hand, the policy network of the Tohoku case was made up of an open network, an *issue network*. While some scholars argue that the policy community type should be considered to increase the efficiency of a disaster response management, others argue that the management system can be more effective with horizontal relationship than vertical because the system becomes more flexible in accepting and coordinating with the relief effort from the outside as Lee and Park [5] address. For the case of Tohoku, while the government was busy downplaying the severity of the nuclear threat and defending its stance, the emergency relief operations by voluntary organizations were notable, and they were busy delivering urgent relief items to the defected population. This study argues that the issue network may be considered as a better network type to increase the efficiency of a disaster response with participation of the private sector.

### B. Contributions

In general, we analyzed the issues on a bureaucratic or hierarchical command control approach to disaster response management and explored the importance of relationship among responding actors in improving the responses. In particular, the current status the local governments have been examined and compared with four different cases, which led to suggesting the importance of the role of a local government with volunteers. This study is meaningful in presenting which concept is relatively important when a disaster response network is formed, and this could give further insight in establishing a more effective network for an unexpected and complex event.

### C. Conclusions

Various network concepts from the policy network theory were investigated to show whether the extent of failures in the disaster management can be improved by connecting the hierarchical and community based open system, which reflects the conclusions from the white paper of the Kobe earthquake, the congressional report of the Katrina, and the report of the National Diet for the Tohoku earthquake. None of the points mentioned from the previous disasters are new or unknown as Rubens [51] argues. In other words, the problems pointed out in the previous Japanese disaster in Kobe can also be found in the hurricane Katrina and the Tohoku earthquake. If the points mentioned from each case are recursive, then it may mean that they are predictable and solvable. The main points on the basis of the previous analyses are: (1) a lack of coordination; (2) a lack of communication; (3) a lack of leadership; (4) a lack of assistance from outside.